# PERIODICALLY PULSED STRONG SQUEEZING

H. Adamyan[1,2], J. Bergou[3], N. Gevorgyan[2], G. Kryuchkyan[1,2]

[1]*Yerevan State University, Alex Manookian 1, 375025, Yerevan, Armenia*
[2]*Institute for Physical Research, National Academy of Sciences, Ashtarak-2, 378410, Armenia*
[3]*University of New York, 695 Park Avenue, New York, NY 10021, USA*

*We report on specific signatures of squeezing for time-modulated light fields. We show that application of periodically-modulated driving fields instead of continuous wave fields drastically improves the degree of quadrature integral squeezing in an optical parametric oscillator. These results particularly allow for applications in time-resolved quantum communication protocols.*

Squeezed states of light play a significant role in quantum information science with continuous variables (CV) [1]. They may serve as a source of CV entanglement since combining two squeezed light beams at a beam splitter creates an entangled two-mode squeezed state. Other interesting applications of squeezed states include, for instance, quantum cryptographic protocols and ultra-high precision measurements. It should be noted that up to now the generation of bright light beam with high degree of squeezing meets serious problems. Indeed, in an optical parametric oscillator (OPO) the integral squeezing reaches only 50% relative to the level of vacuum fluctuations, if the pump intensity is close to the generation threshold.

In this field, a wide variety of squeezing measurements as well as quantum communication applications has been realized in the spectral domain and not in time domain. It should be noted that such spectral measurements (rather than time-dependent) have been performed even for the case of pulsed squeezing experiments [2]. Recently, in addition to these important developments the quantum communication schemes based on time-resolved homodyne measurement of individual pulsed squeezing states as well as an individual quadrature-entangled pulses have been developed [3]. Such single-shot homodyne detection have already been performed in the pioneering experiments on quantum tomography and quantum correlations and also in the nanosecond and picosecond domains [4]. This approach opens possibilities for elaboration of time-resolved quantum information protocols operating in a pulsed regime in addition to the ordinary ones elaborated in spectral domains for continuous waves.

Taking in mind important applications to time-resolved quantum information technology, in this Report we investigate periodically-pulsed squeezed light beams which can be generated in experimentally feasible schemes of OPO under action of pump fields with periodically-varying amplitudes. We stress that application of time-modulated pump fields allows qualitative improve the situation, i.e. to go beyond the fundamental limit 50%. Such improvement of the degree of squeezing is found for both below-and above-threshold regimes.

We consider OPO in two-resonant optical ring cavity (see Fig.1). The Hamiltonian describing this system within the framework of rotating wave approximation and in the interaction picture is

$$H = H_{sys} + \hbar\left(a_L \Gamma_L^+ + a\Gamma^+ + h.c.\right), \qquad (1)$$

where $\Gamma_L$, $\Gamma$ are reservoir operators that create and destroy photons in the loss reservoir coupled to the intracavity modes. The interaction of cavity modes is given by the Hamiltonian

$$H_{sys} = i\hbar f(t)\left(e^{i(\Phi_L - \omega_L t)} a_L^+ - e^{-i(\Phi_L - \omega_L t)} a_L\right)$$
$$+ i\hbar k\left(e^{i\Phi_k} a_L a^{+2} - e^{-i\Phi_k} a_L^+ a^2\right), \quad (2)$$

where $a_L$ and $a$ are the boson operators for cavity modes at the frequencies $\omega_L$ and $\omega_L/2$. The pump mode $a_L$ is driven by an amplitude-modulated external field at the frequency $\omega_L$ with time-periodic, real valued amplitude $f(t+T) = f(t)$. A down-conversion of the pump photons to resonant subharmonic-mode photons at frequency $\omega_L/2$ occurs due to a $\chi^{(2)}$ nonlinearity presented inside the cavity. The constant $ke^{i\Phi_k}$ determines an efficiency of the down-conversion process $\omega_L \to \frac{\omega_L}{2} + \frac{\omega_L}{2}$ in $\chi^{(2)}$ medium.

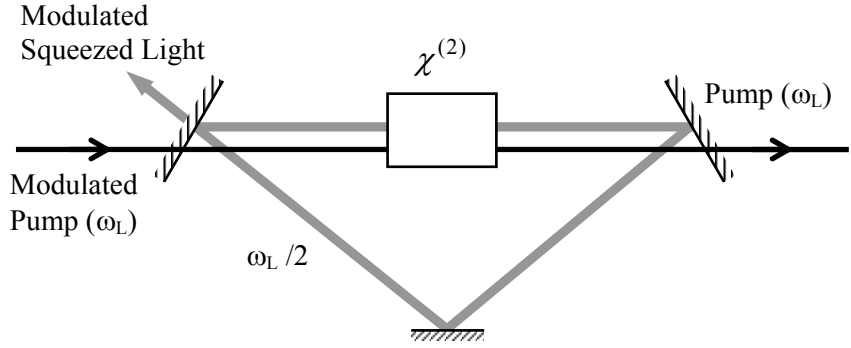

*Fig. 1. The principal scheme of OPO in a cavity that supports the pump mode at frequency $\omega_L$ and subharmonic mode.*

A semiclassical analysis shows that similar to the standard OPO, the considered system also exhibits threshold behavior, which is easily described through the period-averaged pump field amplitude $\overline{f}(t) = \frac{1}{T}\int_0^T f(t)dt$, where a period $T = 2\pi/\delta$, and $\delta$ is a modulation frequency, $\delta \ll \omega_L$. The below-threshold regime with a stable trivial zero-amplitude solution is realized for $\overline{f} < f_{th}$, (where $f_{th} = \gamma \gamma_L/k$ is the threshold value of the pump field amplitude, $\gamma$, $\gamma_L$ are the decay rates of the modes $\omega_L/2$ and $\omega_L$) while above-threshold regime with stable nontrivial solution for mode's amplitudes takes place, if $\overline{f} > f_{th}$.

We focus on the case of OPO driven by pump field with the following continuously, harmonically modulated amplitude $f(t) = f_0 + f_1 \cos(\delta t + \Phi)$. The typical results are demonstrated below for mean photons number as well as for the variance of quadrature amplitude of the generated mode.

In above threshold, $\overline{f} = f_0 > f_{th}$, the mean photon number of the subharmonic $n = \langle a^+ a \rangle$ in over transient regime, $t \gg \gamma^{-1}$, reads as

$$n^{-1}(t) = 2\frac{k^2}{\gamma_L}\int_{-\infty}^{0} \exp\left(2\gamma\tau\left(\frac{\overline{f}}{f_{th}} - 1\right)\right)\exp\left(\frac{2\gamma f_1}{\delta f_{th}}[\sin(\delta(t+\tau)) - \sin(\delta t)]\right)d\tau. \quad (3)$$

This result is illustrated in Fig.2a for the different levels of modulation and for $f_1 = 0$ reaches to the standard result $n = (f_0 - f_{th})/k$ for a continuously pumped OPO in a steady-state.

Let us turn to study the degree of squeezing in the presence of harmonic modulation. As it is well known, the total or integral one-mode squeezing is usually analyzed through the variances $V(\theta)$ of quadrature amplitudes of electromagnetic field $X(\theta) = \frac{1}{\sqrt{2}}(a^+ e^{-i\theta} + a e^{i\theta})$, where $a$, $a^+$ are the boson operators, $V(x) = \langle x^2 \rangle - \langle x \rangle^2$ is a denotation of a variance. It should be noted that an integral squeezing for OPO reaches only 50% relative to the level of vacuum fluctuations, $V \geq 0,5 V_0$, and equality takes place if the pump field intensity close to the generation threshold.

We perform concrete calculations in the frame of stochastic equations for the complex variables corresponding to the operators $a$, $a_L$. The perturbative analysis of quantum fluctuations on the small parameter $k^2/\gamma\gamma_L \ll 1$ of the theory is also used.

The time-dependent variance in the over transient, above-threshold regime is obtained in the following form

$$V(t) = \gamma \int_{-\infty}^{0} \exp\left(-2\int_{\tau}^{0}(\gamma + f(t'+t)k/\gamma_L + \gamma n(t'+t))dt'\right)[1 + 2n(\tau+t)]d\tau, \qquad (4)$$

where the photon number is given by the formula (3). The result for below threshold is obtained from Eq. (4), if $n = 0$.

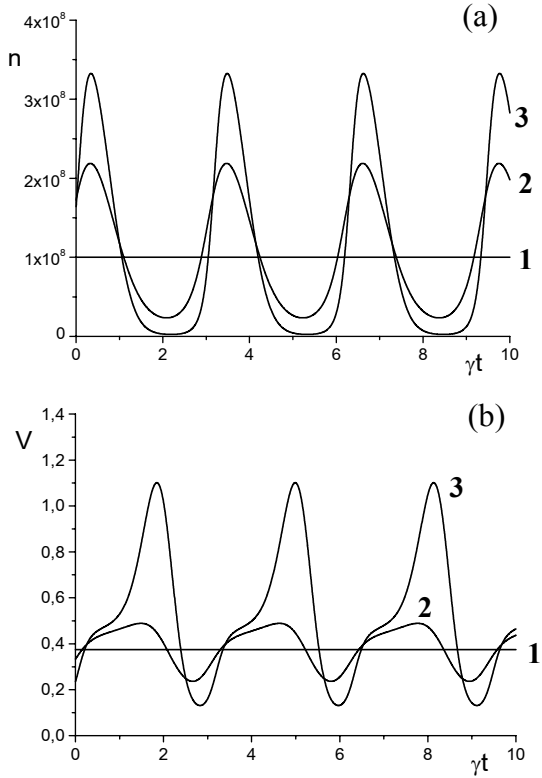

*Fig. 2.* Mean photon number (a) and the variance (b) versus dimensionless time for the parameters:
$k^2/\gamma_L\gamma = 10^{-8}$, $\delta = 2\gamma$, $f_0 = 2\gamma\gamma_L/k = 2f_{th}$;
$f_1 = 0$ (curve 1), $f_1 = 0,75 f_0$ (curve 2), and
$f_1 = 1,5 f_0$ (curve 3).

The variance is seen to show a time-dependent modulation with a period $2\pi/\delta$. The drastic difference between the degree of single-mode squeezing for modulated and stationary dynamics in above-threshold OPO is also clearly seen in Fig. 2b. The stationary variance (curve 1) near the threshold is bounded by the condition $V \geq 0,5 V_0 = 0,25$, while the variance for the case of modulated dynamics can be less than 0,25 for definite time intervals. The minimum values of the variance $V_{\min} = V(t_m)$ at fixed time intervals $t_m = t_0 + 2\pi m/\delta$, $(m = 0,1,2,...)$ synchronized with the modulation period are shown in Fig.3. As it is expected, the degree of squeezing increases with ratio $f_1/f_0$. The production of strong squeezing occurs for the period of modulation comparable with the characteristic time of dissipation, $\delta \approx \gamma$, and disappears for asymptotic cases of slow $(\delta \ll \gamma)$ and fast $(\delta \gg \gamma)$ modulations.

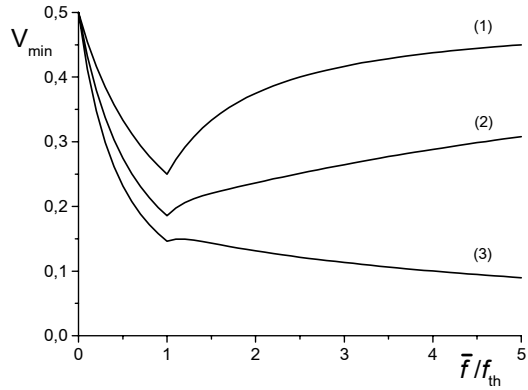

***Fig. 3.*** *The minimum levels of the variance versus* $\bar{f}/f_{th}$ *for three levels of modulation:*
$f_1 = 0$ *(curve 1),* $f_1 = 0.75\bar{f}$ *(curve 2) and* $f_1 = 1.5\bar{f}$ *(curve 3).*
*The parameters are as in the Fig.2.*

The main peculiarity of the scheme proposed in contrast to many other squeezed light sources is that this one operates under the nonstationary condition, because the pump field amplitude is time-dependent. This circumstance has a significant impact on formation of nonclassical light in the presence of dissipation and cavity induced feedback (see, also [5]). We believe that the results obtained can be generalized to a general class of quantum dissipative systems and are applicable to time-resolved quantum information protocols which are now in the stage of development.

**Acknowledgments:** This work was supported by NFSAT - PH 098-02/ CRDF – 12052, ISTC A 823, and ANSEF PS 89-66 grants.